\documentclass[superscriptaddress, amsmath, amssymb, aps,pra,nofootinbib,twocolumn
]{revtex4-2}
\usepackage{graphicx} % Required for inserting images
\usepackage{comment}
\usepackage{dsfont}
\usepackage{color}
\usepackage{bm} % bold math with {\bm ...text...} command

\usepackage[utf8]{inputenc}
\usepackage{amsmath}
\usepackage{amssymb}
\usepackage{float}
\usepackage{braket}
\usepackage{adjustbox}
\usepackage{color}
\usepackage{framed} 
\usepackage{url}
\usepackage[normalem]{ulem}

\usepackage{mathtools} % Include this in your preamble
\usepackage{ragged2e}

\usepackage{float}
\makeatletter
\let\newfloat\newfloat@ltx
\makeatother
\usepackage{algorithm}
\usepackage{algpseudocode} % for \ElsIf and other advanced algorithmic commands
\usepackage{amsmath}
\usepackage{placeins}
\usepackage[colorlinks=true,citecolor=blue,linkcolor=blue,urlcolor=blue]{hyperref}

\begin{document}

\title{Switching Characteristics of Electrically Connected Stochastically Actuated Magnetic Tunnel Junction Nanopillars}

\author{Dairong Chen}
\affiliation{Center for Quantum Phenomena, Department of Physics, New York University, New York, NY 10003 USA}

\author{Ahmed Sidi El Valli}
\affiliation{Center for Quantum Phenomena, Department of Physics, New York University, New York, NY 10003 USA}

\author{Jonathan Z. Sun}
\affiliation{IBM T. J. Watson Research Center, Yorktown Heights, NY 10598, USA}

\author{Flaviano Morone}
\affiliation{Center for Quantum Phenomena, Department of Physics, New York University, New York, NY 10003 USA}

\author{Dries Sels}
\affiliation{Center for Quantum Phenomena, Department of Physics, New York University, New York, NY 10003 USA}
\affiliation{Department of Physics, Boston University, Boston, Massachusetts 02215, USA}
\affiliation{Center for Computational Quantum Physics, Flatiron Institute, New York, NY, USA}

\author{Andrew D. Kent}
\affiliation{Center for Quantum Phenomena, Department of Physics, New York University, New York, NY 10003 USA}

\begin{abstract}
We investigate the stochastic dynamics of nanoscale perpendicular magnetic tunnel junctions (pMTJs) and the correlations that arise when they are electrically coupled. Individual junctions exhibit thermally activated spin-transfer torque switching with transition probabilities that are well described by a Poisson process. When two junctions are connected in parallel, circuit-mediated redistribution of voltages that occurs in real time as the junction resistances change leads to correlated switching behavior. A minimal stochastic model based on single-junction statistical switching properties and Kirchhoff's laws captures the coupled switching probabilities, while a Markov-chain formalism describes nonequilibrium steady states under multi-pulse driving. Further, these circuit-mediated interactions can be mapped onto the parameters of an Ising Hamiltonian, providing an interpretation in terms of effective spin-spin interactions. Our results demonstrate how simple electrical connections can generate Ising-like couplings and tunable stochastic dynamics in nanoscale magnets.
\end{abstract}

\maketitle
\section{Introduction}
There has been growing interest in physics-inspired computing as a means to enhance and complement conventional digital logic based computers. As Landauer emphasized in noting that information is physical~\cite{landauer_information_1991}, the nature of the physical system used to encode and manipulate information is not incidental but fundamental, often enabling entirely new algorithmic possibilities. Quantum computing is a prominent example: algorithms based on qubits, such as Shor’s factoring algorithm~\cite{shor_polynomial-time_1997} and Grover’s search algorithm~\cite{grover_fast_1996}, exploit the quantum nature of information carriers to achieve speedups unattainable with classical bits.

While much of the focus has been on quantum computing, other physical systems are also being explored. An example of such alternative physical systems is found in probabilistic bits (p-bits)~\cite{camsari_stochastic_2017}, often implemented using magnetic tunnel junctions (MTJs) having two well-defined resistance states. A random fluctuation between these states can occur either thermally or through actuation by spin-transfer torques. MTJs can thus function as p-bits, where each bit takes the value 0 or 1 with a certain probability, or they can be viewed as stochastic 3D rotors evolving under nonlinear dynamics~\cite{chen_solving_2025}. Their stochastic and dynamical characteristics make them an interesting candidate for unconventional computing. 

Recent advances in spintronics have highlighted magnetic tunnel junctions (MTJs) as promising building blocks for both Ising machines and neuromorphic computing~\cite{borders_integer_2019,markovic_physics_2020,daniels_neural_2023,misra_probabilistic_2023,shao_probabilistic_2023,nikhar_all--all_2024,si_energy-efficient_2024,cheng_voltage-controlled_2025}. In the context of Ising machines, MTJs have been employed to demonstrate integer factorization with networks of stochastic p-bits~\cite{borders_integer_2019,shao_probabilistic_2023}, to solve a 70-city traveling salesman problem using superparamagnetic devices~\cite{si_energy-efficient_2024}, and to realize scalable architectures with all-to-all reconfigurability~\cite{nikhar_all--all_2024}. On the neuromorphic side, MTJs have been leveraged to implement neural primitives~\cite{daniels_neural_2023}, construct probabilistic circuits for inference~\cite{misra_probabilistic_2023}, and integrated into artificial intelligence pipelines for diffusion-based generative models~\cite{cheng_voltage-controlled_2025}. A perspective on spintronics for non-von Neumann computing is provided in Ref.~\cite{markovic_physics_2020}.

While extensive demonstrations of MTJ-based architectures have been made, in most cases the junctions are treated and driven individually. When realizing an Ising Hamiltonian,
\begin{equation}
    H = -\sum_{\langle i,j \rangle} J_{ij} s_i s_j - \sum_i h_i s_i,
    \label{eq:Ising Hamiltonian}
\end{equation}
the $J_{ij}$ terms are often embedded through external computation rather than arising from actual physical interactions between MTJs. Much less is studied on the emergent dynamics of electrically connected junctions. When multiple MTJs share a circuit, redistribution of voltages and currents can induce correlations in their switching. Such circuit-mediated interactions raise fundamental questions: How strong can the effective coupling be? How can it be controlled? Can the coupled behavior be predicted from the properties of individual junctions? And how do these interactions manifest under nonequilibrium driving?  

Prior work by Talatchian \emph{et al.}~\cite{talatchian_mutual_2021} addressed related questions by demonstrating mutual control of two superparamagnetic MTJs (sMTJs) under DC bias, showing emergent correlations from circuit-mediated effects. In this work, we follow a similar spirit but explore a different regime: we use perpendicular MTJs (pMTJs) that are magnetically stable yet exhibit controlled stochasticity when driven by voltage pulses that induce probabilistic switching events. We experimentally study two pMTJs connected in parallel through a shared circuit and show that the coupled system exhibits effective ferromagnetic- or antiferromagnetic-like interactions depending on the polarity of the applied voltage. Through modeling, we demonstrate that this behavior can be captured within a Poisson-process framework and extend it to arbitrary pulse sequences using a Markov-chain formalism. The Markov approach enables us to predict the non-equilibrium steady-state distributions generated by pulse trains and to map these distributions onto effective exchange interactions between Ising spins.

The motivation for exploring magnetically stable pMTJs lies in their practical and physical advantages. Their stochastic fluctuations can be tuned purely through electrical control—for example, by applying voltage pulses that induce spin-transfer-torque-driven stochastic switching~\cite{Rehm2023,Valli2025}. They also exhibit reduced sensitivity to variations in anisotropy or temperature compared to sMTJs~\cite{Rehm2024,Morshed2023}. Moreover, stable pMTJs are easier to fabricate, compatible with industry-standard CMOS processes, and therefore well suited for scalable implementations. Our results demonstrate that even simple electrical interconnections can mediate effective couplings between magnetically stable pMTJs. Beyond their technological implications, this work establishes a platform for investigating correlated stochastic dynamics and nonequilibrium statistical mechanics in circuit-mediated nanoscale magnetic systems.

The remainder of this paper is organized as follows. Section~\ref{sec:experiment} presents the experimental characterization of individual pMTJs and of two junctions connected in parallel. Section~\ref{sec:modeling} introduces a Poisson-process framework to predict coupled behavior from single-junction properties, extends it to pulse-driven dynamics using a Markov-chain description, and demonstrates how the resulting distributions can be mapped onto the effective exchange coupling of an Ising Hamiltonian.

\section{\label{sec:experiment}Experiments}

\subsection{Single-MTJ Switching \label{sec:single}}

To investigate correlated phenomena in electrically coupled pMTJs, we first characterize the stochastic switching behavior of individual junctions under applied voltage pulses. All measurements are performed at room temperature ($T = 295$~K) on two circular pMTJs with a diameter of 40~nm. These devices are referred to as Device~\textbf{A} and Device~\textbf{B} throughout the study. For simplicity, we denote the parallel (low-resistance) state of a pMTJ as \textbf{0} and the antiparallel (high-resistance) state as \textbf{1}.

\begin{figure*}
    \centering
    \includegraphics[width=1.0\linewidth]{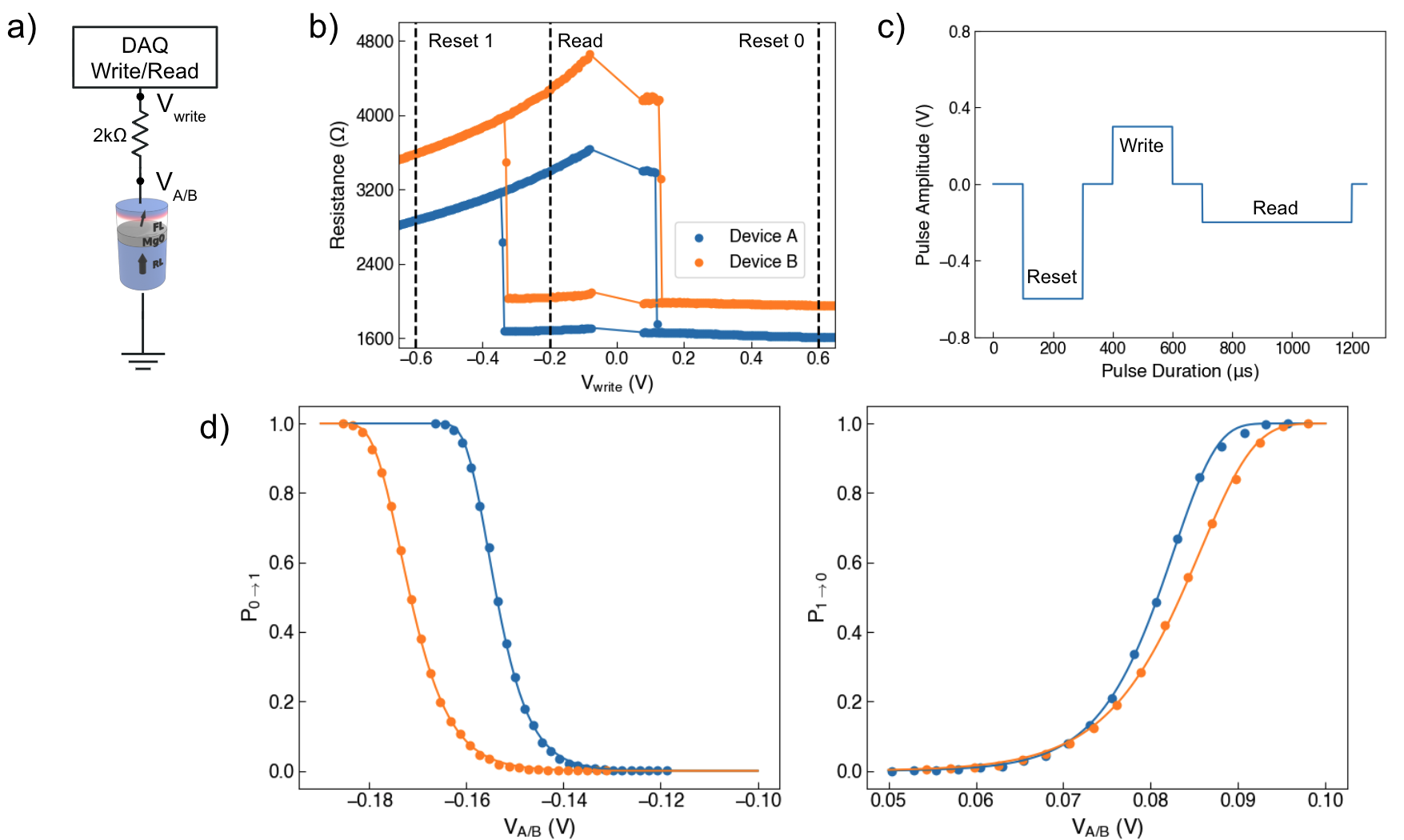}
    \caption{
    (a) Schematic of the experimental setup showing the MTJ connection through a 2~k$\Omega$ resistor. 
    (b) Hysteresis curves for two individual MTJs. Vertical dashed lines indicate reset and read voltages. 
    (c) Example of the waveform used in the experiment. Each sequence consists of a reset pulse to initialize the MTJ state, a write pulse to induce switching, and a read pulse to measure the resulting state. The pulse durations are fixed throughout the experiment. 
    (d) Transition probabilities for the two selected MTJs as a function of $V_A$ or $V_B$. Each point is obtained by repeating the waveform $N=10,000$. The solid lines are fitted sigmoid function as described in Eq.~\ref{eq:sigmoid}, with fitted parameters given by Table.~\ref{tab:params}.}
    \label{fig:singles_summary}
\end{figure*}

The experimental setup shown in Fig.~\ref{fig:singles_summary} was used to measure both the pMTJ quasi-static voltage-resistance hysteresis and the voltage-pulse-dependent switching probabilities. Each device was connected in series with a resistor ($R_0 = 2$~k$\Omega$), and a computer-controlled data acquisition system applied the write voltage ($V_\mathrm{write}$) and measured the voltage across the junctions ($V_\mathrm{A,B}$).

We first measure the voltage-resistance hysteresis loops of both pMTJs by sweeping $V_\mathrm{write}$ and recording the voltage across each junction ($V_\mathrm{A,B}$). The hysteresis loops shown in Fig.~\ref{fig:singles_summary}(b) are then used to determine the reset and read pulse amplitudes. A voltage large enough to set the junction in the \textbf{0} or \textbf{1} state is used for the reset pulses, and voltage small enough to avoid disturbing the magnetic state are used for the read pulse. Switching probabilities are then measured using the reset--write--read pulse sequence shown in Fig.~\ref{fig:singles_summary}(c), in which a 200~$\mu\text{s}$ reset pulse is followed by a variable amplitude 200~$\mu\text{s}$ write pulse and a 500~$\mu\text{s}$ read pulse. This sequence is repeated $N = 10{,}000$ times for each write-pulse amplitude. The resulting switching probabilities (the number of switching events divided by $N$) are plotted in Fig.~\ref{fig:singles_summary}(d) as a function of the voltage across each junction $V_\mathrm{A,B}$, the voltage measured just before switching occurs.

The voltage-dependent transition probability curves form the basis for modeling the behavior of coupled junctions, as discussed later in Section~\ref{sec:model_poisson}. The form of the switching probability in a macrospin model in the thermally activated regime and in the absence of an external magnetic field~\cite{liu_dynamics_2014}, is given by
\begin{equation}
P(V,t) = 1 - \exp \left\{ -\frac{t}{t_0} \exp \left[ -\xi \left( 1 - \frac{V}{V_0} \right) \right] \right\},
\label{eq:sigmoid}
\end{equation}
where $V$ is the pulse amplitude and $t$ is the pulse duration. Here, $1/t_0$ represents the attempt frequency, $V_0$ is the critical voltage, and $\xi = E_B/(kT)$ quantifies the ratio of the energy barrier to magnetization reversal $E_B$ to the thermal energy. In this work, $1/t_0$, $V_0$, and $\xi$ are treated as empirical fitting parameters.

Expression~\ref{eq:sigmoid} can be rewritten to emphasize that the switching probability follows a Poisson process with a voltage-dependent rate. The probability of switching within a time $t$ at an applied voltage $V$ is given by
\begin{equation}
P(V,t) = 1 - \exp[-\Gamma(V) t],
\label{eq:poisson}
\end{equation}
where $\Gamma(V)$ is the voltage-dependent switching rate
\begin{equation}
    \Gamma(V) = \frac{1}{t_0}\exp\left[ -\xi \left(1-\frac{V}{V_0}\right)\right].
    \label{eq:gamma_extract}
\end{equation}
The validity of this Poisson-process assumption as a function of pulse duration is experimentally verified in Appendix~\ref{AppendixA}.

In our experiments, the write-pulse duration is fixed at $t = 200~\mu\text{s}$. 
The switching probabilities shown in Fig.~\ref{fig:singles_summary}(d) are fitted to Eq.~\ref{eq:poisson} 
for each MTJ in both switching directions.  In these fits, the attempt time is fixed at $t_0=1$ ns and $\xi$ and $V_0$ are treated as free fitting parameters.
From these fits, we extract the voltage-dependent switching rates for devices A and B—
$\Gamma_{A,0\rightarrow1}(V)$, $\Gamma_{A,1\rightarrow0}(V)$, 
$\Gamma_{B,0\rightarrow1}(V)$, and $\Gamma_{B,1\rightarrow0}(V)$. 
These experimentally determined rates are then used as inputs to simulate the stochastic dynamics 
of electrically coupled MTJs in 
Section~\ref{sec:model_poisson}. The values of the fitted parameters are listed in Table~\ref{tab:params}.

\begin{table}[t]
\centering
\caption{Fit parameters for the switching probability data of Fig.~\ref{fig:singles_summary}.}
\label{tab:params}

\renewcommand{\arraystretch}{1.2}
\setlength{\tabcolsep}{8pt}

\begin{tabular}{|c|c|c|c|}
\hline
 & $\xi$ & $V_0$ (V) \\
\hline
$\Gamma_{A,0\rightarrow1}$ & 45.4 & $-0.21$ \\
\hline
$\Gamma_{A,1\rightarrow0}$ & 28.9 & \phantom{-}0.14 \\
\hline
$\Gamma_{B,0\rightarrow1}$ & 43.2 & $-0.24$ \\
\hline
$\Gamma_{B,1\rightarrow0}$ & 30.0 & \phantom{-}0.14 \\
\hline
\end{tabular}
\end{table}

\subsection{Coupled MTJs Characteristics}

\begin{figure*}
    \newpage
    \centering
    \includegraphics[width=0.9\linewidth]{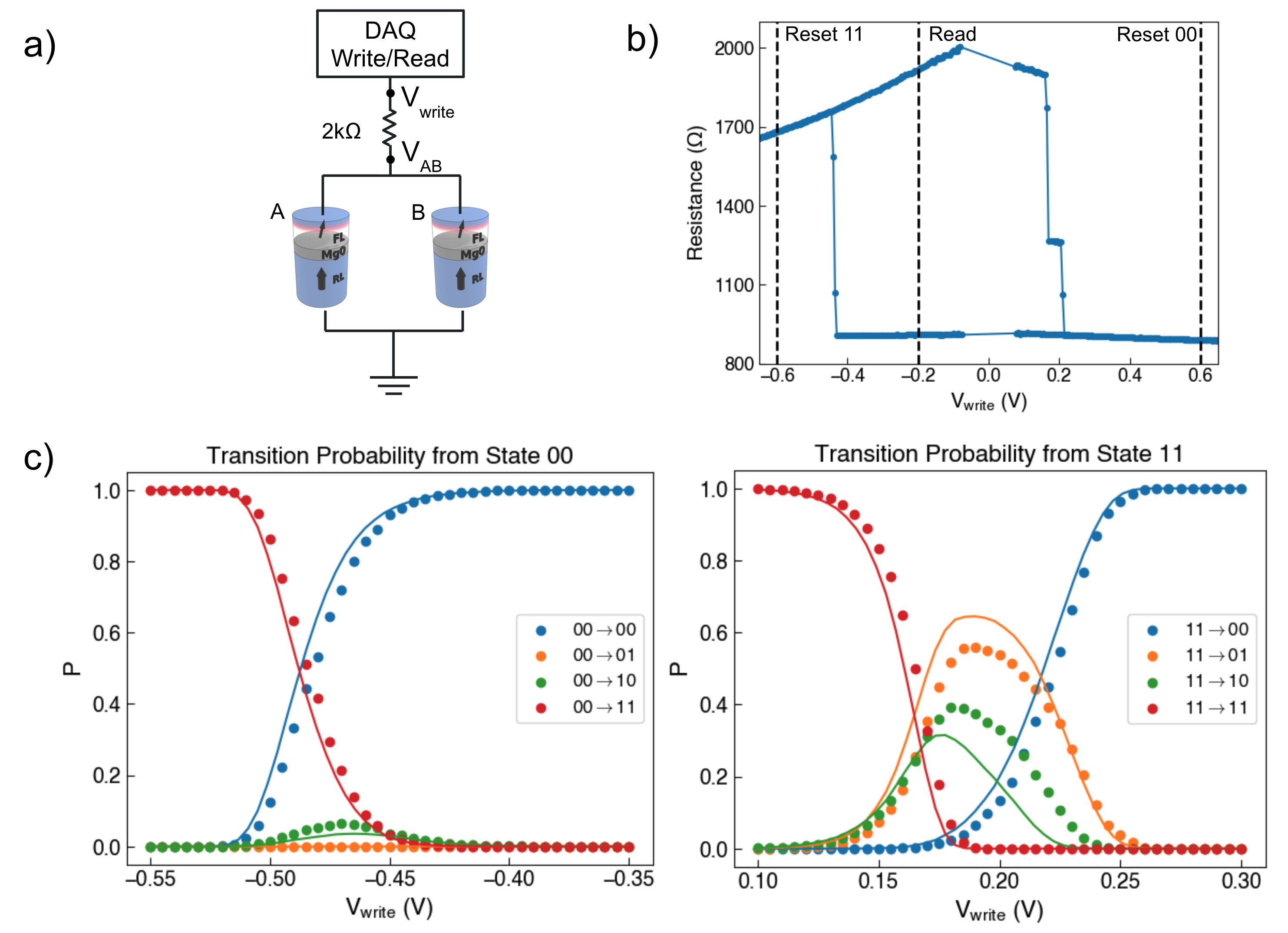}
    \caption{
    (a) Schematic of the experimental setup showing the MTJ connection through a 2~k$\Omega$ resistor. 
    (b) Hysteresis curve of the resistance across the two MTJs connected in parallel. Vertical dashed lines indicate reset and read voltages. A reset voltage $V_\text{rest} = -0.6V$ set both junctions to state 1, and $V_\text{rest} = 0.6V$ set both junctions to state 0.
    (c) Transition probabilities for the two selected MTJs as a function of write pulse amplitude. Each point is obtained by repeating the waveform $N=100,000$. }
    \label{fig:coupled_summary}
\end{figure*}

In this section, we experimentally investigate the behavior of two pMTJs connected in parallel.  Our goal is to test whether circuit-mediated interactions between the junctions lead to correlated switching,  and whether such switching dynamics mimic effective ferromagnetic or antiferromagnetic coupling.  For this purpose, Devices~A and~B are connected in parallel with a series resistor $R_0 = 2\,\text{k}\Omega$, as shown in the schematic in Fig.~\ref{fig:coupled_summary}(a). 

Depending on the state of the two junctions, four different resistance levels can be measured when they are connected in parallel, corresponding to four possible configurations. When both junctions A and B are in low-resistance state 0, denoted as 00, the measured resistance is $R_{00} = 911~\Omega$. When both junctions are in high-resistance state 1, denoted as 11, the resistance is $R_{11} = 1918~\Omega$. The asymmetric states, with A in state 0 and B in state 1 (01) or A in state 1 and B in state 0 (10), have resistance of $R_{01} = 1209~\Omega$ and $R_{01} = 1276~\Omega$ respectively. Since devices A and B have slightly different resistances, the intermediate resistance states 01 and 10 can be distinguished.

The hysteresis loop of the coupled configuration is shown in Fig.~\ref{fig:coupled_summary}(b). As the applied voltage is swept from negative to positive values, both MTJs start from the high resistance state, the state 11. Junction B switches first to the low-resistance state (0), while junction A remains in state 1, resulting in the intermediate resistance state 10. With further increase of the voltage amplitude, junction A also switches to state 0, and the system transitions to the 00 state. When the applied voltage is swept from positive to negative values, the junctions start in the 00 state, followed by an abrupt transition from 00 to 11. This behavior indicates that once one junction switches from state 0 to 1 at negative voltage, the other junction follows immediately, providing evidence of correlated switching. The hysteresis loop also enables determination of the pulse amplitudes required to initialize the junctions in the 00 and 11 states, as well as the readout voltage amplitude, which is chosen to be sufficiently small so as not to perturb the junction states.

To characterize the switching behavior and measure the switching probabilities of the coupled devices, we use the reset–write–readout pulse sequence described previously [Fig.~\ref{fig:singles_summary}(c)], with pulse amplitudes shown in Fig.~\ref{fig:coupled_summary}(b).
The pulses are applied to junctions connected in parallel.  We focus on measuring the transition probabilities from the initial states 00 and 11  to all four possible final states (00, 01, 10, and 11) as a function of the write-pulse amplitude. Each measurement is repeated $N = 100{,}000$ times.

The results of these measurements are shown in Fig.~\ref{fig:coupled_summary}(c). The data show clear voltage-dependent correlations between the two junctions.  When the junctions are initialized to 00, and $V_\text{write} = -0.485\,\text{V}$, the junctions predominantly remain in state 00 or transition into state 11, with probabilities $P_{00} = 0.44$ and $P_{11} = 0.51$. The tendency to end up in a symmetric configuration is indicative of an effective ferromagnetic-like coupling. The correlation arises from the shared circuit: when one junction switches from a low- to a high-resistance state, the voltage across the other junction increases, making it more likely to switch as well.

In contrast, when the junctions are initialized to 11, and \( V_\text{write} = 0.185\,\text{V} \), the pMTJs predominately transition to asymmetric states 01 and 10, with probabilities \( P_{01} = 0.56 \) and \( P_{10} = 0.39 \). This behavior is indicative of an effective antiferromagnetic-like coupling. When one junction switches from a high- to a low-resistance state, the voltage across the other junction decreases, making it less likely to switch and the two junctions end up in opposite configurations.

To fully characterize the coupled dynamics, we also need to measure the transition probabilities between all four possible states. These measurements enable construction of the complete transition matrix, \( T(V) \), for the coupled system---an essential step for the modeling discussed in Sec.~\ref{sec:markov}.
\begin{figure}[h!]
    \centering
    \includegraphics[width=1\linewidth]{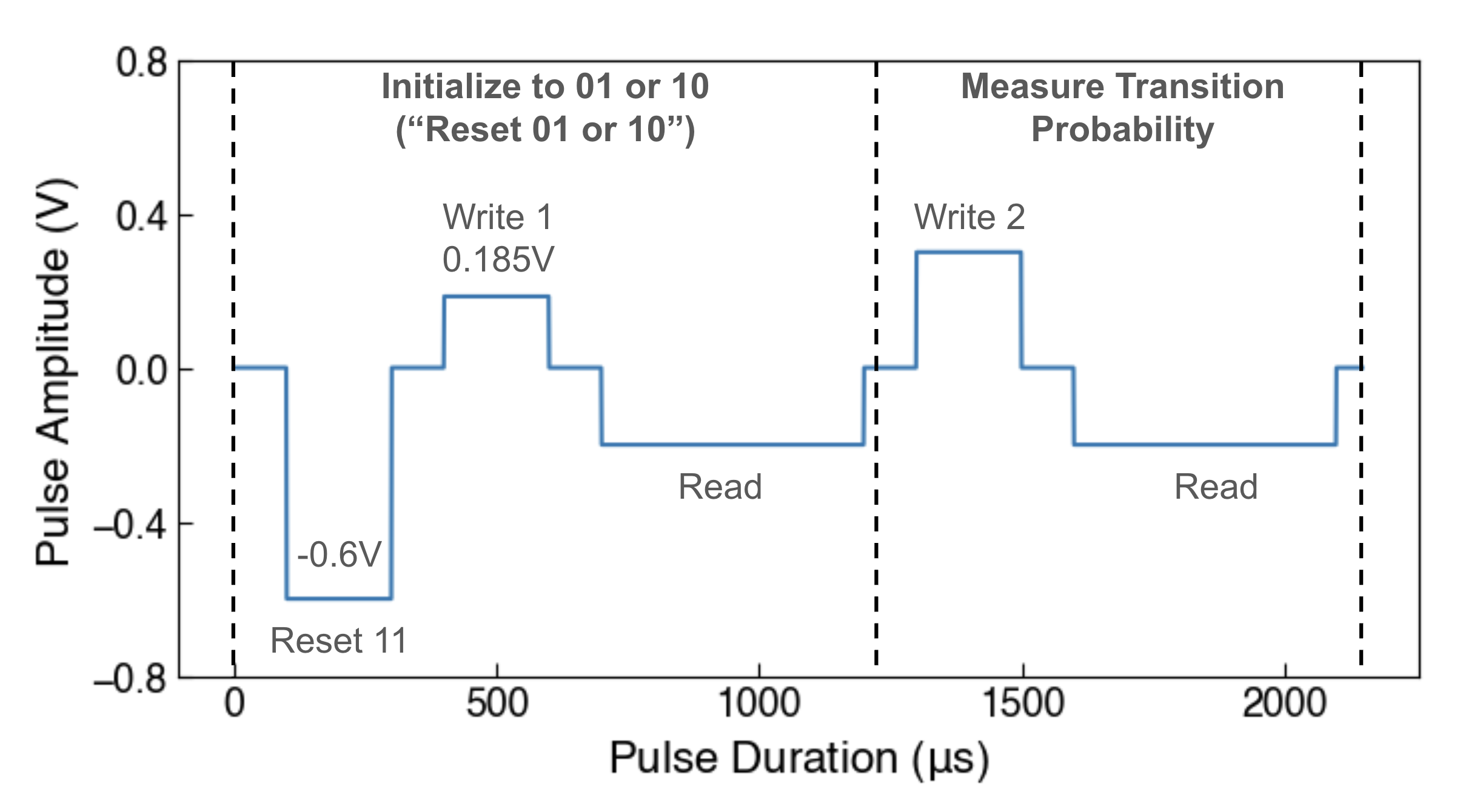}
    \caption{Two-step waveform for measuring the transition probability from state 01 or 10 to other four states. In the first step, the junctions are initialized to either 01 or 10. Then at second step, a variable write pulse is applied to measure the transition probability from 01 or 10 to other four states as a function of applied voltage $V$.}
    \label{fig:two_step_waveform}
\end{figure}

\begin{figure*}
    \newpage
    \centering
    \includegraphics[width=1.0\linewidth]{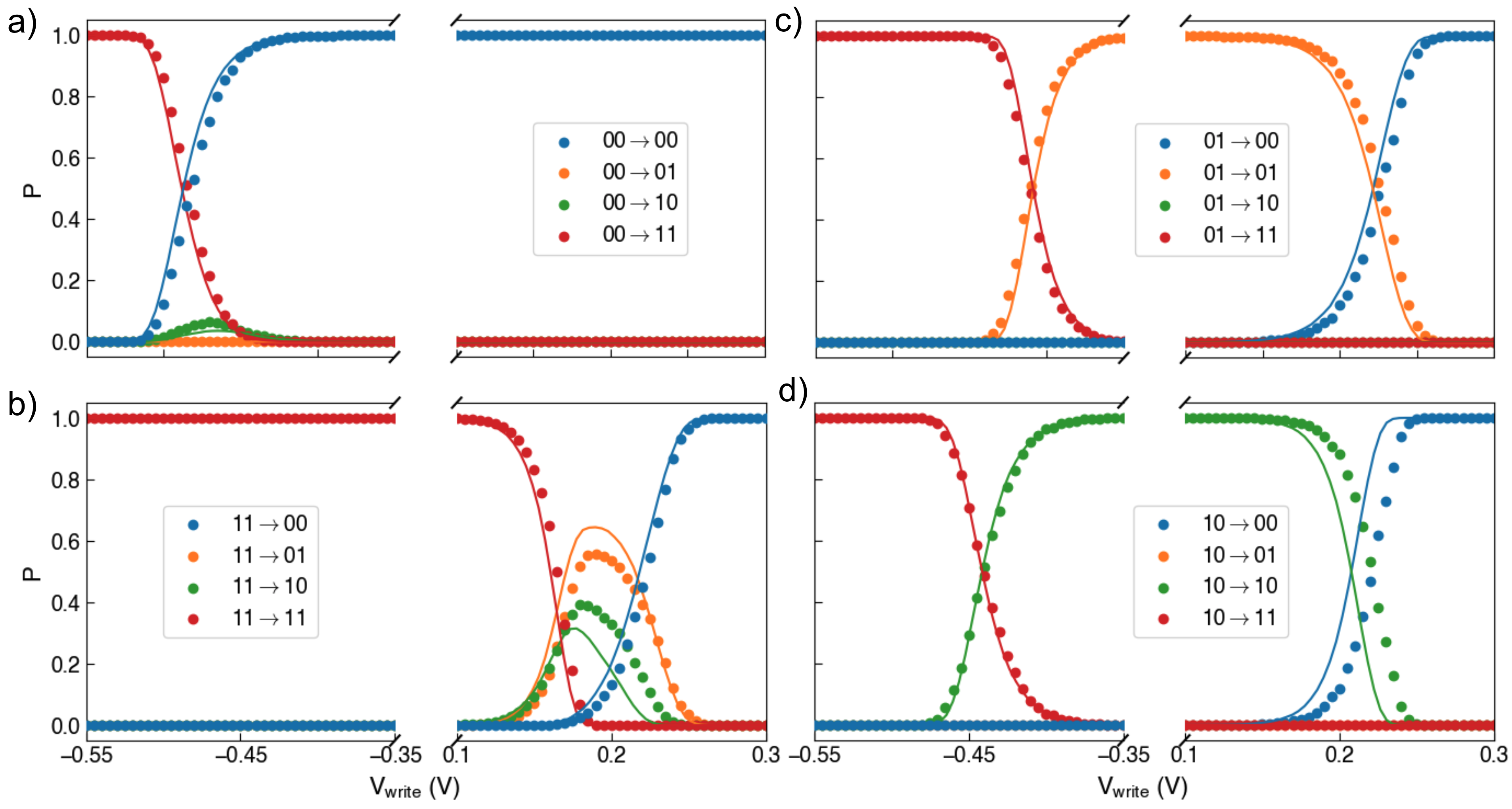}
    \caption{Transition probabilities for all initial states 00, 01, 10,and 11. (a) Transition probabilities from the initial state 00 to all four possible final states. For $V_\text{write} < 0$, the transition probabilities are identical to those shown in Fig.~\ref{fig:coupled_summary}. For $V_\text{write} > 0$, the system remains in state 00 due to the hysteresis of the junctions; positive write voltages do not induce switching from the 0 state. (b) Transition probabilities from the initial state 11 to all four possible final states. For $V_\text{write} > 0$, the transition probabilities are identical to those shown in Fig.~\ref{fig:coupled_summary}. For $V_\text{write} < 0$, the system remains in state 11 due to the hysteresis of the junctions; that is, negative write voltages do not induce switching from the 1 state. (c) Transition probabilities from the initial state 01 to all four possible final states. (d) Transition probabilities from the initial state 10 to all four possible final states.}
    \label{fig:sigmoid_full}
\end{figure*}

This can be done by measuring the transition probabilities from the asymmetric states 01 and 10. To do this a two-step waveform is applied to the pMTJs, as illustrated in Fig.~\ref{fig:two_step_waveform}. This waveform  uses the antiferromagnetic-like behavior observed when the system is initialized in state 11 and driven at \( V_\text{write} = 0.185\,\text{V} \). By selecting only the trials that enter state 01 or 10 after the first write pulse and recording the final state after the second, the transition probabilities from states 01 and 10 to all four possible outcomes are determined. The complete voltage-dependent transition probabilities of the coupled system—from each of the four initial states (00, 01, 10, and 11)—are summarized in Fig.~\ref{fig:sigmoid_full}.

\section{Modeling\label{sec:modeling}}

\subsection{Modeling the Switching Probabilities\label{sec:model_poisson}}

In this section, we develop a modeling framework to simulate the switching behavior of coupled MTJs solely based on properties of individual junctions, and recreate the experimentally observed transition probabilities shown in Fig.~\ref{fig:sigmoid_full}. Such a model is especially valuable for future studies involving more complex systems. As the number of coupled junctions and circuit components increases, experimental characterization becomes increasingly time-consuming and resource-intensive. A predictive model allows us to simulate larger multi-junction architectures for future application designs, and estimate system behavior prior to building the circuit.

We model the switching dynamics of each MTJ as a Poisson process in which the probabilities of transition depends on the instantaneous voltage $V_{AB}$ across the junctions. The coupling between junctions is incorporated through Kirchhoff’s laws: when one junction switches, its resistance changes, and the voltage across the devices is redistributed accordingly. This minimal framework successfully reproduces the correlated probabilistic dynamics observed experimentally in Fig.~\ref{fig:sigmoid_full}, using only the single-junction transition characteristics presented in Fig.~\ref{fig:singles_summary}.

To simulate the stochastic switching behavior of the coupled pMTJs, the pulse duration is discretized into time steps of size \( \Delta t \). At each time step \( t_i \), the applied write pulse voltage and the current state of the junctions determine the instantaneous voltages across the two parallel pMTJs, denoted \( V_{AB,i} \). Using the measured voltage-dependent switching rates \( \Gamma(V) \), the switching probability for each junction at each step is computed as:
\begin{equation}
    P(V_{AB,i}, \Delta t) = 1 - \exp[-\Gamma(V_{AB,i}) \cdot \Delta t].
\end{equation}
These probabilities are then used to update the state of each junction at the next time step. Repeating this procedure over the entire pulse duration \( t_\text{pulse} \) yields a simulated distribution of final states. The detailed implementation of the update rules is provided in the in the algorithm presented in Appendix~\ref{AppendixB}.

The results of the simulation are shown as solid lines in Fig.~\ref{fig:coupled_summary}(d) and Fig.~\ref{fig:sigmoid_full}. While the simulation does not exactly reproduce the experimental measurements, it captures the overall shape, symmetry, and trends of the data. Given the minimal assumptions and simplicity of the model, we consider the level of agreement to be satisfactory for capturing the dynamics of the coupled system.

There are several possible refinements to the model. For example, one could employ a function that more accurately fits the experimental switching curves in Fig.~\ref{fig:singles_summary}(d), or incorporate the voltage dependence of the pMTJ resistance, which decreases in the antiparallel state at high bias. Such effects could be included in future iterations of the model.

\subsection{Sequential Pulse Dynamics and Markov Chain Model\label{sec:markov}}

In this section, we extend our analysis from the dynamics within a single pulse to the evolution of the system under sequences of pulses. Because the pulse durations used in our experiments are relatively long, the system is expected to relax between successive pulses, allowing each pulse to be treated as independent. Under this assumption, the evolution of the junctions under a sequence of pulses can be described as a discrete-time Markov process.

The Markov-chain perspective allows us to consider more general pulse trains with varying amplitudes and durations, provided that the Markov assumption remains valid. When a given pulse train is applied repeatedly, the system evolves toward a steady-state distribution characterized by the probabilities $(P_{00}, P_{01}, P_{10}, P_{11})$. The steady-state distribution is also independent of the initial states that we started. By designing the sequence of voltage pulses—each characterized by its own transition matrix—we can tailor the system’s output distribution using purely electrical control.

To simulate the system’s steady-state behavior under a sequence of pulses, we first construct the transition matrix $T$ as a function of the applied voltage $V$. The transition matrix depends on the pulse duration $t$; however, since all pulses in our experiments have a fixed duration of $t = 200~\mu\text{s}$, we restrict our analysis to the voltage dependence alone. The effect of a single voltage pulse \(V\) is represented by a transition matrix \(T(V)\), defined as
\begin{equation}
\scalebox{0.8}{$
T(V) =
\begin{bmatrix}
P_{00 \to 00}(V) & P_{00 \to 01}(V) & P_{00 \to 10}(V) & P_{00 \to 11}(V) \\
P_{01 \to 00}(V) & P_{01 \to 01}(V) & P_{01 \to 10}(V) & P_{01 \to 11}(V) \\
P_{10 \to 00}(V) & P_{10 \to 01}(V) & P_{10 \to 10}(V) & P_{10 \to 11}(V) \\
P_{11 \to 00}(V) & P_{11 \to 01}(V) & P_{11 \to 10}(V) & P_{11 \to 11}(V)
\end{bmatrix}
$}.
\end{equation}
Each element \(P_{ij \to i'j'}(V)\) gives the probability of transitioning from an initial state \(ij\) to a final state \(i'j'\) under an applied voltage \(V\). These transition probabilities are determined from the experimental data shown in Fig.~\ref{fig:sigmoid_full}.  

To model a sequence of pulses with different amplitudes or durations, we multiply the corresponding transition matrices in the order the pulses are applied. The resulting matrix product captures the cumulative evolution of the system, enabling prediction of its steady-state probability distribution after any desired pulse sequence. For a pulse train consisting of \(n\) pulses, denoted \(V_\text{train} = [V_1, V_2, \ldots, V_n]\), the overall transition matrix is given by
\begin{equation}
    T(V_\text{train}) = T(V_1) \, T(V_2) \cdots T(V_n).
\end{equation}
When this pulse train $V_\text{train}$ is applied repeatedly, the system evolves toward a stationary, non-equilibrium steady state. The steady-state distribution is derived by solving for the normalized eigenvector of \(T^\top(V_\text{train})\) associated with eigenvalue one.

\begin{figure*}[t]
    \centering
    \includegraphics[width=0.8\textwidth]{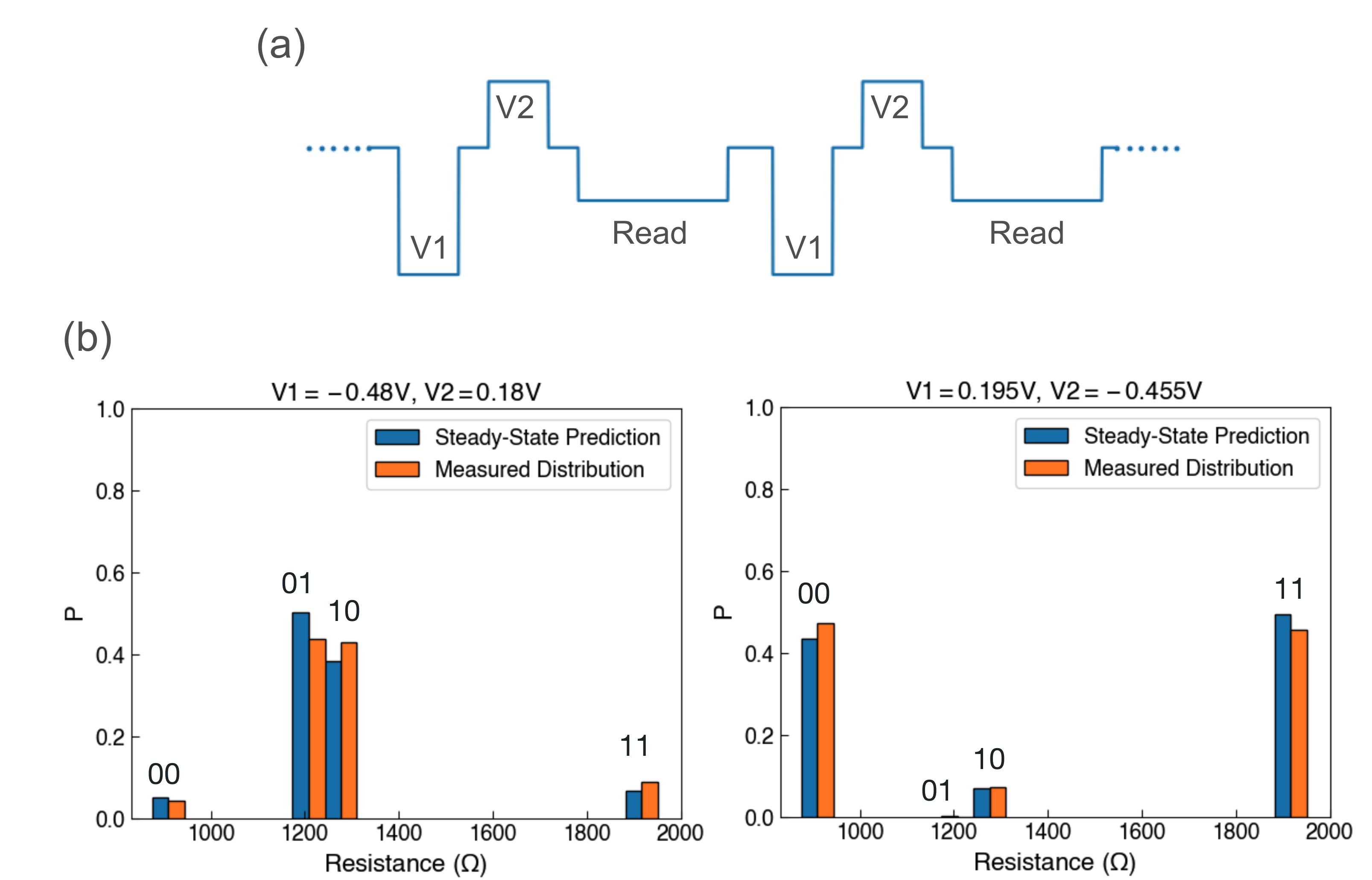}
    \caption{(a) Schematic of the waveform used to measure the steady-state distribution. The same pulse sequence, $V_1$–$V_2$–Read, is applied $N = 10{,}000$ times.(b) Comparison between experimentally measured and theoretically predicted steady-state distributions for alternating pulse sequences. Two cases are shown: one with $V_1 = -0.48\,\text{V}$ and $V_2 = 0.18\,\text{V}$, and another with $V_1 = 0.195\,\text{V}$ and $V_2 = -0.455\,\text{V}$.}
    \label{fig:steady_states}
\end{figure*}

We experimentally tested the Markov-chain model by applying a pulse train composed of two voltages, $V_\text{train} = [V_1, V_2]$, to junctions A and B connected in parallel. The pulse sequence was repeated $N = 10{,}000$ times, and a schematic of the waveform is shown in Fig.~\ref{fig:steady_states}(a). The resulting state distributions were compared with the predicted steady states, as shown in Fig.~\ref{fig:steady_states}(b). The measured distributions show good agreement with the theoretical predictions, indicating that the Markov-chain framework accurately captures the pulse-sequence-driven dynamics of the system.

Given the agreement between the Markov-chain framework and the experimental data, we use the model to predict the steady-state distributions accessible through arbitrary pulse trains $V_\text{train} = [V_1, V_2]$ and $V_\text{train} = [V_1, V_2, V_3]$, where the voltage values are constrained by the resolution of the experimentally applied pulses. The resulting steady-state probabilities are shown in Fig.~\ref{fig:J_h1_h3}(c) and Fig.~\ref{fig:J_h1_h3}(e). This approach can be readily extended to longer pulse sequences involving four or more pulses.

These results demonstrate that the probabilistic behavior of a coupled pMTJ system can be tuned through purely electrical control, without modifying the underlying circuit. Mapping pulse sequences to steady-state distributions provides a versatile framework for programmable stochastic control.

\section{Application: Physical Couplings for Ising Machines}
In the broader context of Ising-machine research using MTJs, our architecture can be viewed as a physical realization of spin–spin couplings. The corresponding Ising Hamiltonian, which describes pairwise spin interactions and local fields, is given by
\begin{equation*}
    H = -\sum_{\langle i,j \rangle} J_{ij} s_i s_j - \sum_i h_i s_i,
\end{equation*}
where \( s_i \in \{-1, 1\} \) are spin variables, \( J_{ij} \) denotes the coupling strength between spins \(i\) and \(j\), and \( h_i \) represents local fields.

In our system, each pMTJ acts as an effective Ising spin: the low-resistance parallel state (\(0\)) is denoted \(s=-1\), and the high-resistance antiparallel state (\(1\)) is denoted \(s=+1\). In our system, when two MTJs are electrically coupled, their joint behavior under applied voltage pulses can be described by an effective Hamiltonian,
\begin{equation}
    H(s_1, s_2) = -J s_1 s_2 - h_1 s_1 - h_2 s_2 + C,
\end{equation}
where \(J\) represents the effective coupling between spins \(s_1\) and \(s_2\), \(h_1\) and \(h_2\) are local field terms, and \(C\) is a constant offset. For example, the energy of both spins in the low resistant sate, $E_{00}$, would given by:
\begin{equation*}
  E_{00} = H(s_1 = -1, s_2 = -1) = -J  + h_1 + h_2 + C,
\end{equation*}
and one can work out the cases for any $E_{ij}$.

In most Ising machine architectures, the system is assumed to obey the Boltzmann distribution. The probability of getting a certain state should follow $P_{ij} = e^{-\beta E_{ij}}/Z$, with inverse temperature $\beta$ and partition function $Z$. We choose to work in dimensionless units by setting $\beta=1$. Substituting the energy $E_{ij}$, one can derive the relation between the probabilities of states, $(P_{00}, P_{01}, P_{10}, P_{11})$, and the effective parameters in the Ising Hamiltonian $(J, h_1, h_2)$:
\begin{equation}
    \begin{pmatrix}
        J \\ h_1 \\ h_2 \\ C'
    \end{pmatrix}
    = \frac{1}{4}
    \begin{pmatrix}
        1 & -1 & -1 & 1 \\ 
        -1 & -1 & 1 & 1 \\ 
        -1 & 1 & -1 & 1 \\ 
        -1 & -1 & -1 & -1 
    \end{pmatrix}
    \begin{pmatrix}
        \ln P_{00} \\ \ln P_{01} \\ \ln P_{10} \\ \ln P_{11}
    \end{pmatrix},
    \label{eq:mapping}
\end{equation}
where \(C' = C + \ln Z\).  

\begin{figure*}[t]
    \centering
    \includegraphics[width=0.8\linewidth]{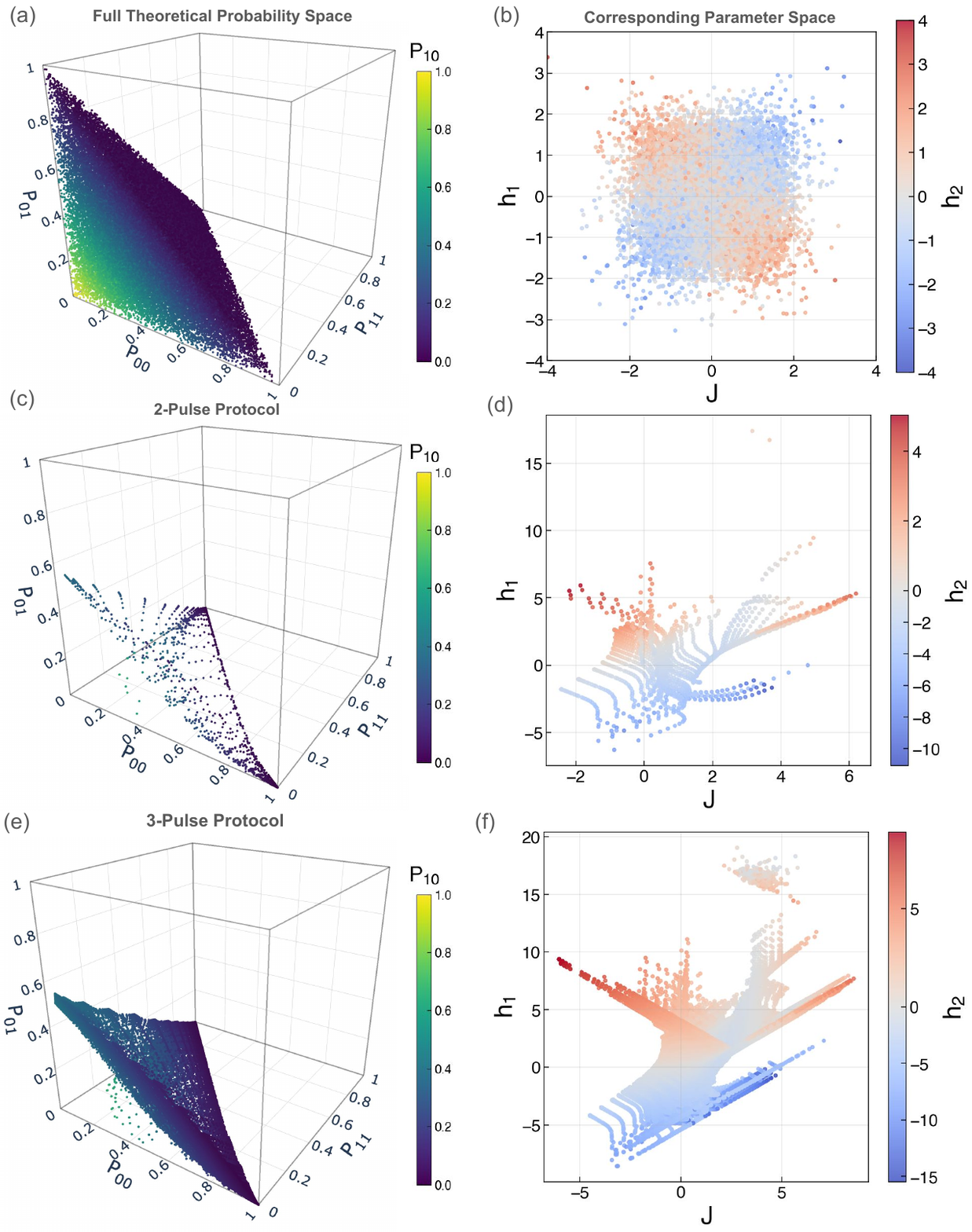}
    \caption{
(a) Full allowed theoretical probability space of $(P_{00}, P_{01}, P_{10}, P_{11})$. The axes correspond to $P_{00}$ (x-axis), $P_{01}$ (y-axis), and $P_{11}$ (z-axis), while the color encodes $P_{10}$. The allowed probability space forms a simplex with unit-length edges.
(b) Corresponding parameter spaces $(J, h_1, h_2)$ obtained using Eq.~(\ref{eq:mapping}) from the probability distributions shown in panel (a).
(c) Steady-state distributions attainable using a two-pulse sequence $V_\text{train} = [V_1, V_2]$. The accessible distributions are derived from the experimentally measured switching probabilities of the specific pair of MTJs (Devices A and B) used in this experiment.
(d) Corresponding parameter spaces $(J, h_1, h_2)$ obtained using Eq.~(\ref{eq:mapping}) from the probability distributions shown in panel (c).
(e) Steady-state distributions attainable using a three-pulse sequence $V_\text{train} = [V_1, V_2, V_3]$. The accessible distributions are derived from the experimentally measured switching probabilities of the specific pair of MTJs (Devices A and B) used in this experiment.
(f) Corresponding parameter spaces $(J, h_1, h_2)$ obtained using Eq.~(\ref{eq:mapping}) from the probability distributions shown in panel (e).
}
    \label{fig:J_h1_h3}
\end{figure*}

This formulation provides a direct mapping between the experimentally measured probability distribution and the corresponding effective coupling and local field parameters, provided that all $P_{ij} > 0$. To examine the structure of this mapping, we first sample the full space of theoretically allowed probability distributions, shown in Fig.~\ref{fig:J_h1_h3}(a), and map them to the $(J, h_1, h_2)$ parameter space in Fig.~\ref{fig:J_h1_h3}(b). The mapping is not surjective: for $J < 0$, only parameter sets with $h_1$ and $h_2$ having the same sign are accessible, whereas for $J > 0$, only configurations with opposite-sign local fields appear. Developing alternative mappings that overcome these constraints is an interesting direction for future work; here we adopt the present approach because it is the most straightforward and physically intuitive.

Using the Markov-chain framework introduced in Sec.~\ref{sec:markov}, we compute the steady states attainable under pulse trains consisting of two and three voltage pulses, based on the experimentally measured switching probabilities in Fig.~\ref{fig:sigmoid_full}. The resulting steady-state distributions are shown in Fig.~\ref{fig:J_h1_h3}(c) and Fig.~\ref{fig:J_h1_h3}(e), and their corresponding mappings to the effective parameters $(J, h_1, h_2)$ are shown in Fig.~\ref{fig:J_h1_h3}(d) and Fig.~\ref{fig:J_h1_h3}(f). These results demonstrate that tailored pulse sequences can access a variety of effective couplings and local fields, thereby enabling electrically tunable interactions between junctions.

The peculiar structure observed in the $(J, h_1, h_2)$ space arises primarily because the accessible probability distributions of the coupled junctions lie on a curved surface. We attribute this behavior in part to the asymmetry between the two MTJs; if the devices were identical, the distributions would lie strictly on the triangular surface defined by $P_{01} = P_{10}$. In experiment, however, the two junctions exhibit slightly different switching characteristics,
breaking this symmetry. As a result, the predicted steady-state manifold deviates from the flat triangular plane and becomes a gently curved surface in probability space.

Overall, this work establishes a concrete mapping between the probability distributions generated by two electrically coupled MTJs and the effective parameters $(J, h_1, h_2)$ of an Ising Hamiltonian, where the distributions are controlled through sequences of electrical pulses. Expanding the accessible probability space in future implementations can be done by incorporating additional circuit elements such as resistors or nonlinear components. The framework presented here provides a conceptual foundation for engineering electrically programmable interactions in stochastic MTJ systems, and offers a starting point for exploring more complex circuit-mediated coupling schemes in future device architectures.

\section{\label{sec:conclusion}Conclusion}

We have demonstrated that two MTJs connected in parallel through a series resistor exhibit circuit-mediated interactions that give rise to emergent correlated switching behavior. Depending on the polarity of the applied voltage, the system shows preferences for either aligned states (00 and 11), resembling ferromagnetic coupling, or anti-aligned states (01 and 10), resembling antiferromagnetic coupling. These behaviors arise purely from the redistribution of voltage across the parallel network due to Kirchhoff's laws, without any direct magnetic interaction between the junctions.

The switching dynamics of each junction can be modeled as a Poisson process, using the voltage-dependent switching probabilities extracted from single-MTJ characteristics. This modeling framework, when combined with basic circuit analysis, reproduces the correlated behavior of the coupled system with good agreement to experimental results. Furthermore, when extended to multi-pulse driving, the system dynamics can be described using a Markov chain formalism, enabling prediction of the steady-state distribution under arbitrary pulse sequences.

The ability to program interactions through electrical control alone---without modifying the physical hardware---opens new possibilities for reconfigurable stochastic systems and spin-based computing architectures. From a physics standpoint, our results highlight how nonequilibrium steady states can emerge from simple driven sequence of pulses, offering a new platform for exploring the dynamics of coupled junctions.

\subsection*{Acknowledgements} The work at NYU is supported by the Office of Naval Research (ONR) under Award
No. N00014-23-1-2771. The authors thank Karlo de Leon for helpful discussions and insightful comments.

\appendix
\section{Testing the Poisson-Process Assumption}
\label{AppendixA}

In Sections~\ref{sec:single} and~\ref{sec:model_poisson}, we assumed that each MTJ follows a Poisson process, where the switching probability is given by Eq.~\ref{eq:poisson}. In the experiments presented in Fig.~\ref{fig:singles_summary}, we fixed the pulse duration at $t = 200~\mu\text{s}$ and measured the switching probability as a function of the applied voltage $V$. The resulting data were well described by the fitted curves, as shown in Fig.~\ref{fig:singles_summary}(d).

\begin{figure*}[b]
    \centering
    \includegraphics[width=0.8\textwidth]{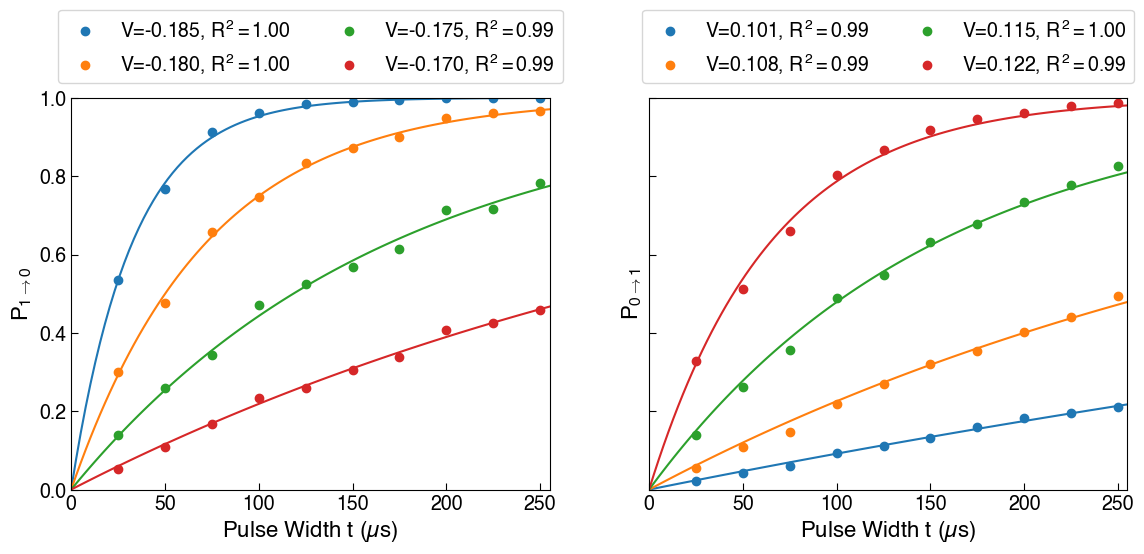}
    \caption{Switching probability as a function of write-pulse duration for a single perpendicular magnetic tunnel junction (pMTJ) at several fixed pulse amplitudes. Symbols show experimentally measured switching probabilities for transitions $1 \rightarrow 0$ (left panel) and $0 \rightarrow 1$ (right panel). Solid lines are fits to the Poisson-process model $P(V,t)=1-\exp[-\Gamma(V)t]$. The excellent agreement ($R^2 \approx 0.99$--$1.00$) validates the Poisson-process description used to extract switching rates for the modeling of coupled junctions and multi-pulse dynamics.}
    \label{fig:test_poisson}
\end{figure*}

To test the validity of Eq.~\ref{eq:poisson} over time, we performed additional measurements in which the write-pulse duration was systematically varied while keeping the voltage amplitude fixed. The goal is to determine whether the measured switching probabilities $P(V, t)$ follow the expected exponential dependence $P(V, t) = 1 - \exp[-\Gamma(V)t]$. The experimental result are shown in Fig.~\ref{fig:test_poisson}. We fitted Eq.~\ref{eq:poisson} to the experimental data and found good $R^2$ agreement. This result further confirms the validity of treating the system as a Poisson process. 

\section{Algorithm used to model electrically coupled MTJs}
\label{AppendixB}
\begin{algorithm}
\caption{Stochastic Flip Update for Coupled MTJs} 
\begin{algorithmic}[1]
\State \textbf{Input:} Initial states $A_i,B_i \in \{0,1\}$; $V_{\text{write}}$; $\Delta t$
\State \textbf{Constants:} $R_A(0),R_A(1),R_B(0),R_B(1)$; $R_0$; $\Gamma_A$, $\Gamma_B$
\State Compute $R_{\text{AB}} \gets$ parallel resistance of $R_A(A_i)$ and $R_B(B_i)$
\State Compute $V_{\text{AB}} \gets V_{\text{write}} \cdot \frac{R_{\text{AB}}}{R_{\text{AB}} + R_0}$

\If{$A_i = 0$ and $B_i = 0$}
    \If{$V_{\text{write}} \geq 0$}
        \State No switching: $A_{i+1} \gets A_i$, $B_{i+1} \gets B_i$
    \Else
        \State $P_A \gets 1 - \exp(-\Gamma_A(V_{\text{AB}}) \cdot \Delta t)$
        \State $P_B \gets 1 - \exp(-\Gamma_B(V_{\text{AB}}) \cdot \Delta t)$
        \State $A_{i+1} \gets$ flip with probability $P_A$
        \State $B_{i+1} \gets$ flip with probability $P_B$  
    \EndIf

\ElsIf{$A_i = 1$ and $B_i = 1$}
    \If{$V_{\text{write}} > 0$}
        \State $P_A \gets 1 - \exp(-\Gamma_A(V_{\text{AB}}) \cdot \Delta t)$
        \State $P_B \gets 1 - \exp(-\Gamma_B(V_{\text{AB}}) \cdot \Delta t)$
        \State $A_{i+1} \gets$ flip with probability $P_A$
        \State $B_{i+1} \gets$ flip with probability $P_B$
    \Else
        \State No switching: $A_{i+1} \gets A_i$, $B_{i+1} \gets B_i$
    \EndIf

\ElsIf{$A_i = 0$ and $B_i = 1$}
    \If{$V_{\text{write}} > 0$}
        \State $P_B \gets 1 - \exp(-\Gamma_B(V_{\text{AB}}) \cdot \Delta t)$
        \State $A_{i+1} \gets A_i$
        \State $B_{i+1} \gets$ flip with probability $P_B$
    \ElsIf{$V_{\text{write}} < 0$}
        \State $P_A \gets 1 - \exp(-\Gamma_A(V_{\text{AB}}) \cdot \Delta t)$
        \State $A_{i+1} \gets$ flip with probability $P_A$
        \State $B_{i+1} \gets B_i$
    \Else
        \State No switching: $A_{i+1} \gets A_i$, $B_{i+1} \gets B_i$
    \EndIf

\ElsIf{$A_i = 1$ and $B_i = 0$}
    \If{$V_{\text{write}} > 0$}
        \State $P_A \gets 1 - \exp(-\Gamma_A(V_{\text{AB}}) \cdot \Delta t)$
        \State $A_{i+1} \gets$ flip with probability $P_A$
        \State $B_{i+1} \gets B_i$
    \ElsIf{$V_{\text{write}} < 0$}
        \State $P_B \gets 1 - \exp(-\Gamma_B(V_{\text{AB}}) \cdot \Delta t)$
        \State $A_{i+1} \gets A_i$
        \State $B_{i+1} \gets$ flip with probability $P_B$
    \Else
        \State No switching: $A_{i+1} \gets A_i$, $B_{i+1} \gets B_i$
    \EndIf
\EndIf

\State \textbf{Output:} Updated states $A_{i+1}$, $B_{i+1}$
\end{algorithmic}
\end{algorithm}

\FloatBarrier
\bibliographystyle{apsrev4-2}
\bibliography{mybib}
\end{document}